\documentclass[12pt,a4paper]{iopart}
\usepackage{amsfonts,iopams,cite}
\usepackage{mathrsfs}
\usepackage{graphicx}
\usepackage{dcolumn}  
\usepackage{psfrag}
\usepackage[usenames]{color}
\newcommand{\ket}[1]{\left\vert#1\right\rangle}
\newcommand{\bra}[1]{\left\langle#1\right\vert}

\newcommand{\beq}{\begin{equation}}
\newcommand{\eeq}{\end{equation}}

\renewcommand{\min}[0]{\mathrm{min}}
\renewcommand{\emph}[1]{{\it #1}}

\newcommand{\ignore}[1]{}

\begin{document}

\title{Thermometry Precision in Strongly Correlated Ultracold Lattice Gases}

\author{M. Mehboudi}
\address{Departament de F\'{\i}sica, Universitat
Aut\`{o}noma de Barcelona, E-08193 Bellaterra, Spain}
\ead{mohammad.mehboudi@uab.cat}

\author{M. Moreno-Cardoner}
\address{Centre for Theoretical Atomic, Molecular and Optical Physics Queen's University, Belfast BT7 1NN, United Kingdom}

\author{G. De Chiara}
\address{Centre for Theoretical Atomic, Molecular and Optical Physics Queen's University, Belfast BT7 1NN, United Kingdom}

\author{A. Sanpera}
\address{ICREA, Instituci\'o Catalana de Recerca i Estudis Avan\c{c}ats, E-08010 Barcelona, Spain}
\address{Departament de F\'{i}sica, Universitat Aut\`{o}noma de Barcelona, E-08193 Bellaterra, Spain}

\date{\today}

\begin{abstract}
The precise knowledge of the temperature of an ultracold lattice gas simulating a strongly correlated system is a question of both, fundamental and technological importance. Here, we address such question by combining tools from quantum metrology together with the study of the quantum correlations embedded in the system at finite temperatures. Within this frame we examine the spin-$1/2$ XY chain, first estimating, by means of the quantum Fisher information, the lowest attainable bound on the temperature precision. We then address the estimation of the temperature of the sample from the analysis of correlations using a quantum non demolishing Faraday spectroscopy method. 
Remarkably, our results show that the collective quantum correlations can become optimal observables to accurately estimate the temperature of our model in a given range of temperatures.

\end{abstract}



\section{Introduction}\label{sec:introduction}
Ultracold atomic samples are considered to be, nowadays, one of the most promising setups for implementing quantum simulators of condensed matter \cite{Bloch2008, Lewenstein2012, Salomon2012}. Such promise has been reinforced by several breakthroughs which include, among others, the celebrated Mott insulator to superfluid quantum phase transition for bosons \cite{Greiner2002}, as well as recent simulations of antiferromagnetic spin chains with both, bosonic \cite{Simon2011} and fermionic \cite{Tarruell2012} ultracold atomic gases. 

At zero temperature, the emergence of a new order in a strongly correlated system is signalled by the presence of quantum correlations at all length scales. At finite temperature, however, such emergence fades gradually away due to the presence of thermal fluctuations. As a result, for low dimensional systems, critical points signalling quantum phase transitions often broaden into ``critical'' regions.  Those regions still separate different phases which keep track of their ground state correlations. Hence, the transition between those phases might appear as smooth crossovers \cite{Mermin-Wagner1,Mermin-Wagner2}, nonetheless carrying a footprint of the quantum phase transition occurring at zero temperature. In view of these facts, finite temperature quantum correlations could be used as a method for thermometry.
Achieving low enough temperatures to simulate strongly correlated systems and other exotic phenomena has been considered as the guiding principle of ultracold lattice physics. Difficulties to reach such regimes arise first  from the inability to measure the temperature on such systems which is a necessary step in order to cross 
the frontier towards strongly correlated ultracold atoms \cite{McKay2010}.

As it is well known in quantum metrology, the quantum Cram{\'e}r-Rao bound \cite{lloyd,paris,braunstein_caves} settles a limit on the precision of the estimation of a given parameter. If the parameter to
be estimated is temperature and the system is in thermal equilibrium, the Cram\'{e}r-Rao
bound for a single shot yields a relation of the form $\Delta T \Delta H \geq T^{2}$ being $H$ the Hamiltonian governing the system and  where we have settled the Boltzman constant $k_{B}=1$ \cite{Jing, Zanardi, Correa14}. This relation indicates that the minimal error in temperature estimation of a thermal sample is realized by a projective measurement on its energy eigenbasis.  In general, such type of measurements in ultracold lattice gases is not accessible. Instead, information about quantum phases and temperature is usually obtained from momentum and density distributions or from density-density (or spin-spin) correlations. These quantities can be extracted by using destructive methods such as time of flight imaging (the latter via the study of noise correlations \cite{Folling2005}) or in-situ imaging, for instance using single site addressability \cite{Sherson2010, Bark2010}. Despite their huge relevance, these methods might suffer limitations in certain occasions, due to their destructive character. For instance, in order to study spin-spin correlations in currently available setups for single site imaging, one needs to remove all particles from one of the two spin components. In this sense, quantum non demolition (QND) methods can provide clear advantages \cite{Eckert2008}.  The quantum Faraday spectroscopy is a minimally disturbing matter-light interface that maps collective atomic quantum correlations into light quadrature fluctuations, the latter observable to be measured by homodyne detection.  Here, we adapt this method to estimate the temperature of a strongly correlated system simulated by an atomic lattice gas. Furthermore, to assess the reliability of our method for precision thermometry, we compare the signal-to-noise ratio obtained from the measurement of collective atomic correlations with the minimal possible error provided by the quantum Cram{\'e}r-Rao bound. Our results show that the measurement of collective quantum correlations can become 
optimal for temperature estimation in some integrable models.

The paper is organized as follows. In Section 2, we briefly review the basic properties of the spin-$1/2$ XY chain in a transverse field, both at zero and finite temperatures. Unlike the majority of quantum spin models, the XY model can be exactly solved by means of a Jordan-Wigner transformation mapping it onto a system of non-interacting fermions and giving access to the full energy spectrum \cite{Lieb61}. In Section 3, we focus on the quantum metrology aspects of the problem. To this aim, we derive first a closed form of the quantum Fisher information (QFI) as a function of the temperature for the whole phase diagram. This, in turn, provides the minimal error on the temperature estimation when performing an optimal measurement.  Section 4 reviews the basic concepts describing the QND Faraday spectroscopy, while Section 5 is devoted to the analysis of quantum correlations at finite temperatures with this method.  We evaluate, for the whole phase diagram of the model, the signal-to-noise ratio, $T/\Delta T$, obtained with a Faraday interface. As we will show later, the thermal sensitivity of a given quantum phase strongly depends on the temperature of the sample. Remarkably, our results support the suitability of collective quantum correlations as optimal observables for quantum thermometry of strongly correlated systems in many cases. In Section 6 we conclude and present some open questions. 

\section{ The XY model}
The spin-$1/2$ XY chain in a transverse field (including the Ising and isotropic XX models as particular cases) is an exactly solvable model, and as such, it can be used as a  prototype to understand the interplay between quantum and thermal fluctuations. The Hamiltonian governing the system can be written as:
\beq
H=-J\sum_{i=1}^{N}\left[\frac{1+\gamma}{2}  \sigma^{x}_{i}\sigma^{x}_{i+1}+\frac{1-\gamma}{2} \sigma^{y}_{i}\sigma^{y}_{i+1}\right]-h\sum_{i=1}^N\sigma_{i}^{z}
\label{XYmodel}
\eeq
where $\sigma_i^\alpha$ are the usual Pauli matrices at site $i$, $-1\leq \gamma \leq 1$ is the parameter that sets the XY anisotropy ($\gamma = \pm 1$ and $\gamma = 0$ for Ising and XX models respectively), $h$ is the transverse magnetic field and $N$ is the number of sites of the chain. The coupling constant $J$ can be positive (ferromagnet) or negative (antiferromagnet). Throughout this paper, we will consider only the ferromagnetic case $J>0$. However, equivalent results can be straightforwardly derived for the antiferromagnetic case $J<0$.   
For simplicity, we consider here periodic boundary conditions with an even number of sites, but the results can be easily extended to an odd number of sites or an open chain. However, for large enough chains, one expects such variations not to influence the results \cite{Lieb61}. 

The Hamiltonian (\ref{XYmodel}) can be easily diagonalized by mapping it onto a non-interacting fermionic model that provides the full energy spectrum. As it is well known \cite{Lieb61, Mikeska} the non-interacting fermionic representation of the XY model is obtained by means of the Jordan-Wigner transformation, followed by a unitary Bogoliubov transformation in the quasi-momentum space, yielding the separable Hamiltonian (up to a constant):
\begin{equation}
H=\sum_{k} \epsilon_{k} \gamma_{k}^{\dag}\gamma_{k},
\label{XYJordan}
\end{equation}
and the energy dispersion relation 
\begin{equation}
\epsilon_k=2J \sqrt{\left(\cos k- h/J\right)^2+\left(\gamma\sin k \right)^2},
\label{positive-energy}
\end{equation}
being $k$ the quasi-momentum, $k=\frac{\pi}{N}(2j+1)$, and $j=-N/2,\dots,N/2-1$. The sign of this energy is arbitrary. Choosing a positive value corresponds to the particle-hole picture for the fermionic quasiparticles,  which are defined for $k\in (0,\pi)$ by the following Bogoliubov transformation:  
\beq
\gamma^\dagger_{\pm k} = \cos \theta_k c^\dagger_{\pm k} \pm  i\sin \theta_k c_{\mp k}.
\eeq
Here, $\cos (2\theta_k) = \gamma \sin k / (\cos k - h/J)$  for $\theta_k \in (0,\pi/2)$, and $c^\dagger_k$ are the Fourier transform of the on-site fermionic operators that directly relate to the spin operators via the Jordan-Wigner transformation 
\beq
c^\dagger_l =  \sigma^+_l \prod_{l'<l} \sigma_{l'}^z\;, \;\;\;
c_l = \sigma^-_l \prod_{l'<l} \sigma_{l'}^z \;\;.
\eeq
The ground state of the system corresponds to the vacuum of the Bogoliubov quasiparticles, and excitations are obtained with creation operators acting on the vacuum. The energy gap between the ground state and the continuum of excited states is thus given by $\Delta E = \min_{k} (\epsilon_k)$.

\begin{figure}[h!]
\centering
\includegraphics[width=12cm]{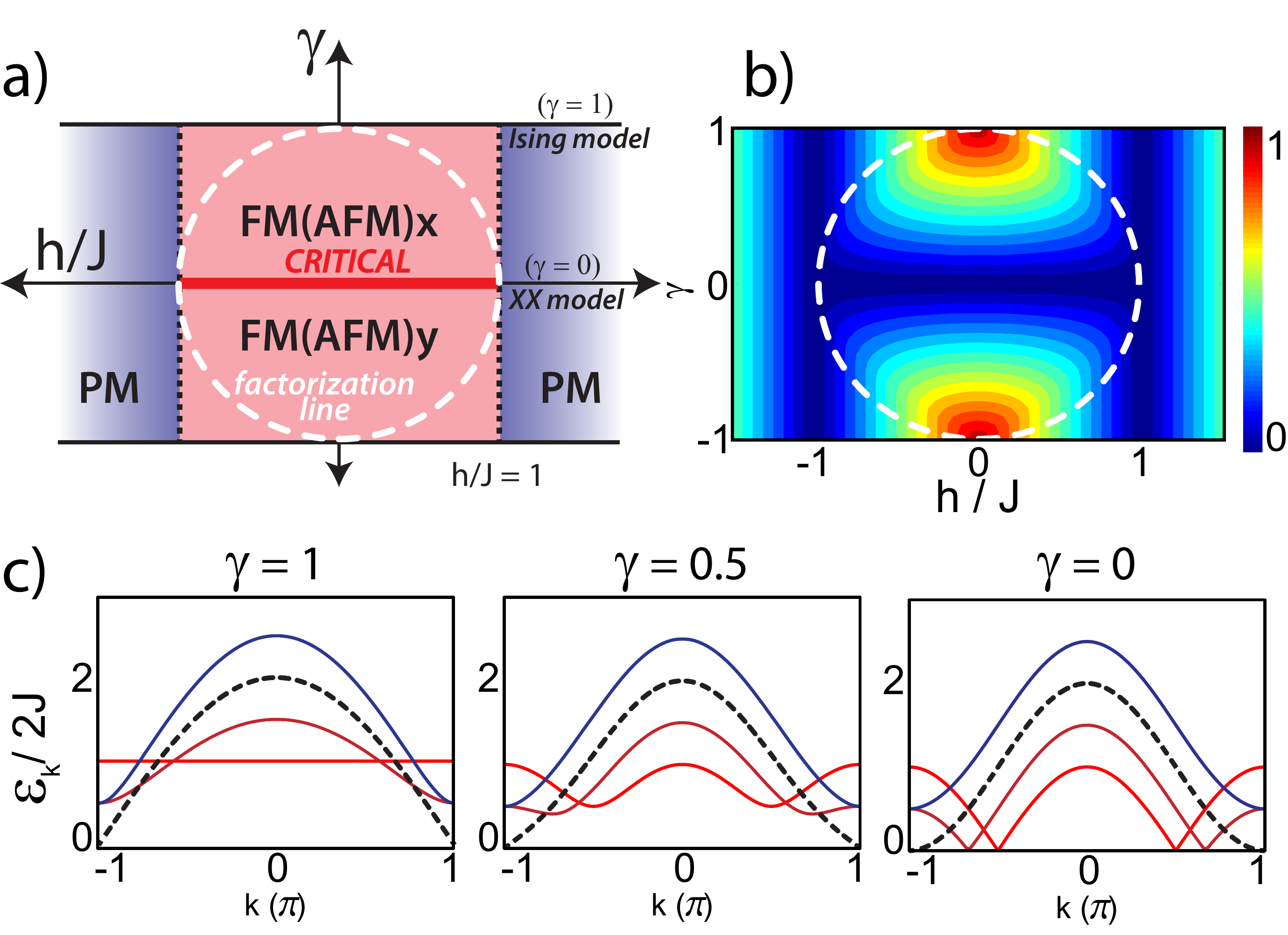}
\caption{(Color online) (a) Sketch of the phase diagram at zero temperature for the XY model. The $\gamma=0$ and $\gamma =1$ lines correspond to the isotropic XX  and Ising models respectively. FM(AFM) denote phases with quasi long-range ferro(antiferro)-magnetic order along the x- and y-axis for $\gamma>0$ and $\gamma<0$, respectively. PM is the paramagnetic phase. There are second order phase transitions at $h/J=\pm 1$ (Ising transition) and at $\gamma = 0$ (anisotropy transition). The dashed line denotes the factorization line for this model. (b) Energy gap $\Delta E$ to the continuum of excited states (in units of $2J$). The energy spectrum is always gapped except at the critical point $h/J=\pm 1$ and at the critical phase $\gamma = 0$, $|h/J|<1$. (c) Energy dispersion relation for different values of the anisotropy parameter $\gamma$.  FM phases are displayed in red ($h/J=0$) and dark red ($0<h/J<1$).  PM phases are displayed in blue ($h/J>1$). Critical points are displayed by the dashed back line ($h/J=1$).}
\label{Figure1}
\end{figure}

Note that the Hamiltonian is symmetric under the exchange $h\leftrightarrow -h$ (by $k\leftrightarrow \pi/2 -k$) and under  $\gamma \leftrightarrow -\gamma$ (by $\sigma_x \leftrightarrow \sigma_y$). A sketch of the phase diagram at zero temperature, together with the energy gap $\Delta E$ and the energy dispersion relation are displayed in Fig.~\ref{Figure1}. The system is always gapped, i.e. $\Delta E >0$, except for the quantum critical lines occurring at $h/J = \pm 1$ (Ising transitions), which separate the paramagnetic phases (PM) from the  ferromagnetic (FM) ones  (or antiferromagnetic if $J<0$) and for $\gamma=0$ and $|h/J|\leq 1 $, corresponding to the critical phase in the XX model (anisotropic transition). Moreover, Heisenberg systems with general anisotropies exhibit, for particular values of the couplings, a ground state which is doubly degenerated and which is factorizable as a product of on-site localized wave-functions \cite{Kurman82, Giampaolo}. In the XY model, for each value of $\gamma$, this product ground state corresponds to an external transverse field $h / J=\pm \sqrt{1-\gamma^2}$, which is depicted by a dashed line in the phase diagram of Fig.~\ref{Figure1}(a).

In the thermodynamic limit (large $N$), the system in thermal equilibrium at a given temperature $T$ can be described by the density matrix in the macrocanonical ensemble (we set $k_B=1$): 
\begin{equation}
\varrho(\gamma, h/J,T/J)=\frac{ e^{-H(\gamma, h,J)/T}} {\mathcal{Z}}= \bigotimes_{k} \varrho^k(\gamma, h/J,T/J),
\label{XYthermal}
\end{equation}
where ${\mathcal{Z}}$ denotes the partition function of the system. For compactness of notation we write from now on $\varrho^{(k)}(\gamma, h/J,T/J)$ simply as $\varrho^{(k)}(T)$. Since the Hamiltonian (\ref{XYJordan}) is separable, the density matrix can be directly written as a tensor product of the density matrices 
associated to each quasiparticle mode $k$. These quasiparticles obey fermionic commutation relations, and thus
\begin{equation}
\varrho^k(T)=\frac{\ket{0}_k\bra{0}+\mathrm{e}^{-\epsilon_k / T}\ket{1}_k\bra{1}}{1+\mathrm{e}^{-\epsilon_k/T}},
\label{Thermal-Product-State}
\end{equation}
where $\ket{0}_k$ ($\ket{1}_k$) denotes an empty (occupied) quasiparticle state $k$. We take the above expression as the starting point to study correlations at finite temperatures.

Finally, let us remark that the XY model can be realistically implemented in experiments. In particular, the isotropic XX model directly maps onto a system of hard-core bosons and it has been experimentally realized with cold atoms in optical lattices \cite{Paredes2004}, while the Ising model has been also engineered with a similar system \cite{Simon2011}.  Moreover, other models that can be implemented with cold atoms, as the bond-charge Hubbard model, directly map onto the XY model \cite{Roncaglia2010}.\\

\section {Optimal strategy: lowest bound on the temperature error}
Consider the state of our strongly correlated system given by $\varrho(T)$. This state depends on the value of the temperature $T$, which is unknown and that we want to estimate. In general, if a quantum state depends on an unknown parameter $\theta$ that we want to estimate, the typical strategy is to choose an unbiased estimator 
$\hat{\theta}$ for which $\langle\hat{\theta}\rangle=\theta$ and repeat the estimation $\nu$ times.  The standard deviation of this estimator, i.e. $\Delta\hat{\theta}=\sqrt{\mathrm{Var}(\hat{\theta})}$, quantifies the error on estimation of $\theta$. 
The quantum Cram\'{e}r-Rao bound sets a lower bound on this error as follows~\cite{lloyd,paris}:
\begin{eqnarray}
(\Delta\hat{\theta})^2\geq\frac{1}{\nu\mathcal{F}(\theta)} .
\label{CRB}
\end{eqnarray}
The factor $\nu$ just follows from the central limite theorem, and $\mathcal{F}(\theta)$ is the quantum Fisher information (QFI) associated to the parameter $\theta$, which is given by:
\beq
{\mathcal{F}(\theta)}=\mathrm{Tr}[\varrho_{\theta}\mathrm{\Lambda}_{\theta}^2],
\eeq
where the symmetric logarithmic derivative, $\Lambda_{\theta}$, is defined as
\beq
\partial_{\theta}\varrho_{\theta}=\frac{\varrho_{\theta}\mathrm{\Lambda}_{\theta}+\mathrm{\Lambda}_{\theta}\varrho_{\theta}}{2}.
\eeq
For temperature estimation on a Gibbs state $\varrho(T)$, the QFI is explicitly given by \cite{Haupt, Zanardi}:
\begin{equation}
{\mathcal{F}(T,\varrho(T))}=\frac{\Delta H^{2}}{T^{4}},
\label{QFIT}
\end{equation}
where $\Delta H^2\equiv \Tr( H^2\varrho(T))- [\Tr(H \varrho(T))]^2$.\ignore{and we have used the identity $\langle H\rangle=T^2 \partial_T \ln \mathcal{Z}$.}\ignore{ For Gibbs states this can be connected to the specific heat \cite{Zanardi}.} Maximizing the quantum Fisher information is hence equivalent to maximize the variance of the Hamiltonian. Introducing the thermal energy as $T$ (note that $k_B$ is set to one), it is possible to express the quantum Cram{\'e}r-Rao bound in the form of an uncertainty relation \cite{Correa14,Zanardi}, that for a single shot reads
\beq
\Delta H\frac{\Delta T}{T^2} \geq 1,
\label{uncertainty}
\eeq
or equivalently, $\Delta H \Delta \beta \geq 1$. This provides a very useful insight to understand how the thermal energy, the energy spectrum of the Hamiltonian and the error on the temperature determination come into play. Indeed, according to (\ref{CRB}), quantum states having a larger QFI can be estimated with a smaller error. As a figure of merit, we define the thermal sensitity as the value of the bound obtained for a single shot ($\nu=1$).  In this way, we withdraw the statistical dependence on the number of times the sample is probed. 

In general, finding  the corresponding QFI of a system is a very difficult task, and different bounds on the QFI that are easier to evaluate, as suggested in \cite{Alipour-Mehboudi,Escher,Alipour-Rezakhani}. In the temperature estimation of a strongly correlated thermal state, the  difficulty arises in the calculation of its intricate energy spectrum, and, in general, it is not possible to derive a closed expression for the QFI. However, such calculation becomes straightforward for the XY model due to the simple structure of a thermal state which corresponds to a product state in the fermionic representation (\ref{XYthermal}), (because of the fact that the Hamiltonian itself (eq. 2) is separable in this representation). From this it follows trivially that the QFI, being linked to the uncertainty of the Hamiltonian, has to be additive, which allows us to express $\mathcal{F}(T,\varrho(T))$ as the sum of the QFI $\mathcal{F}(T,\varrho^{k}(T))$ of each individual mode $k$, i.e.
\beq
\mathcal{F}(T,\varrho(T))=\frac{(\Delta H)^2}{T^4}= \sum_{k} \mathcal{F}(T,\varrho^k(T))
= \sum_{k} \left(\frac{\epsilon_k}{T^{2}}\right)^2  n_k (1-n_k)
\label{Eq:QFI_Allk}
\eeq
being $n_k=(1+e^{\epsilon_k/T})^{-1}$ the Fermi-Dirac distribution of the quasiparticles. 

Using (\ref{uncertainty}) and (\ref{Eq:QFI_Allk}), the upper bound on the signal-to-noise-ratio is given by
\beq
(T / \Delta T)^2_{\mathrm{CRB}}  = T^2 \mathcal{F}(T,\varrho(T)) = \sum_{k} \left(\frac{\epsilon_k}{T}\right)^2  n_k (1-n_k).
\label{Eq:stn}
\eeq

\begin{figure}[h!]
\centering
\includegraphics[width=14cm]{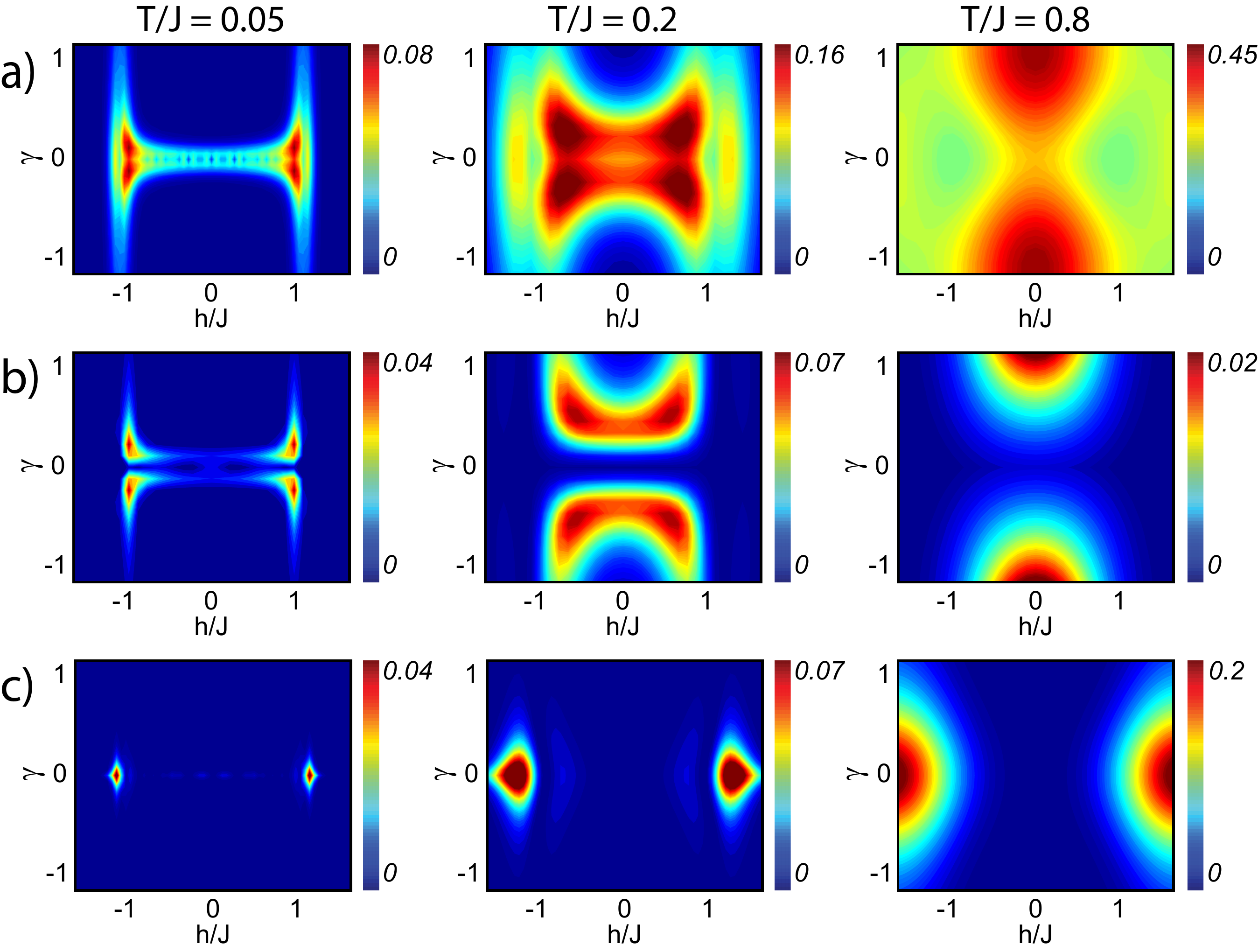}
\caption{(Color online). a) Optimal signal-to-noise ratio, $(T/\Delta T)^{2}_{\mathrm{CRB}}$, where $\Delta T$ denotes the temperature uncertainty given by the Cram\'er-Rao bound when assuming the optimal measurement strategy, plotted as a function of the Hamiltonian parameters and for different values of $T/J$. At very low $T$, the thermal sensitivity is larger close to the critical points, whereas when increasing $T$, the maximum gradually shifts to the Ising and $h=0$ point. (b) and (c) Signal-to-noise ratio, $(T/\Delta T)^{2}_\mathrm{F}$, estimated for the Faraday interface for the two mean values of the observables $(J_x-\langle J_x \rangle)^2$ and $J_z$, respectively. The $\mathrm{Var} (J_x)$ is more sensitive in the FM phase, whereas $\langle J_z \rangle$ works better in the PM phase. All the figures are normalized by the number of atoms ($N=50$ here). Also notice that the color scales are different in each plot.}
\label{Figure2}
\end{figure}

In the top panels of Fig.~\ref{Figure2}, we display this upper bound, normalized by the total number of sites $N$, for the whole phase diagram at different temperatures. For finite $T$, this quantity scales linearly with $N$. For very small temperatures, e.g. $T / J=0.05$, the QFI becomes  noticeable only close to the critical lines. This is not surprising, since for a gapless system, excitations to the lowest part of the energy spectrum will be created no matter how small the temperature is. Thus, as the uncertainty in energy of the state grows, so does the QFI, and accordingly the state becomes very sensitive to thermal fluctuations. In contrast, for a gapped phase, if $T \ll \Delta E$, the probability of creating excitations remains low. In such cases, the energy remains well defined, yielding a vanishing value of the QFI and correspondingly a large error in temperature estimation. On the other hand, for large enough values of the temperature, i.e. $T \geq \Delta E$, different modes become excited, and other regions of the phase diagram become more sensitive and optimal for thermometry. In fact, for a given value of $T$, the accurate estimation of the sample  temperature depends on the energy spectrum but also on the density of states (DOS), as they play a crucial role in the QFI expression (\ref{Eq:QFI_Allk}). This can be clearly seen in Fig.~\ref{Figure2}, where the value of the optimal signal-to-noise ratio $(T /\Delta T)^2_{\mathrm{CRB}}$ is also displayed for $T/J=0.2$ and $T/J=0.8$. The more sensitive regions of the phase diagram are now clearly different than the ``zero temperature transition points'', i.e. $h=\pm1$ and $\gamma=0$ for $|h/J|<1$. In the same figure, in the middle and bottom panels, we display the signal-to-noise ratio obtained from measuring collective correlations that we will analyze in Section 5.

Finally, the behavior of the QFI or thermal sensitivity with temperature is explicitly shown for some particular cases in Fig.~\ref{Figure5} (solid lines). After displaying a maximum at certain value of $T/J$, this quantity decreases again as the state tends to be maximally disordered.  Indeed, at very large temperatures ($\beta\rightarrow 0$), and despite the variance $\Delta H$ is maximum and the error $\Delta \beta$ is minimum, the signal-to-noise ratio $T/\Delta T = \beta / \Delta \beta$ will tend to zero.

\section{Quantum Faraday Spectroscopy}
Here, we briefly review a quantum non-demolition scheme for measuring quantum correlations in ultracold atomic lattices. The method is based on a light-matter interface \cite{Kupriyanov} employing the quantum Faraday effect. It was adapted to determine quantum phases of  strongly correlated systems in optical lattice systems in \cite{Eckert07,Eckert2008}. The scheme is extremely versatile and can detect superfluidity, superlattice ordering and itinerant magnetism for fermionic and bosonic lattice gases \cite{roscilde09,rogers14}. It also allows to reconstruct the phase diagram of non-trivial spin chain models \cite{dechiara11a,dechiara11b} and to engineer quantum correlations by suitable post-selection \cite{hauke13}. In the following we review the basics of the scheme but we point the reader to the previous references for more details. 

\begin{figure}[h!]
\centering
\includegraphics[width=10cm]{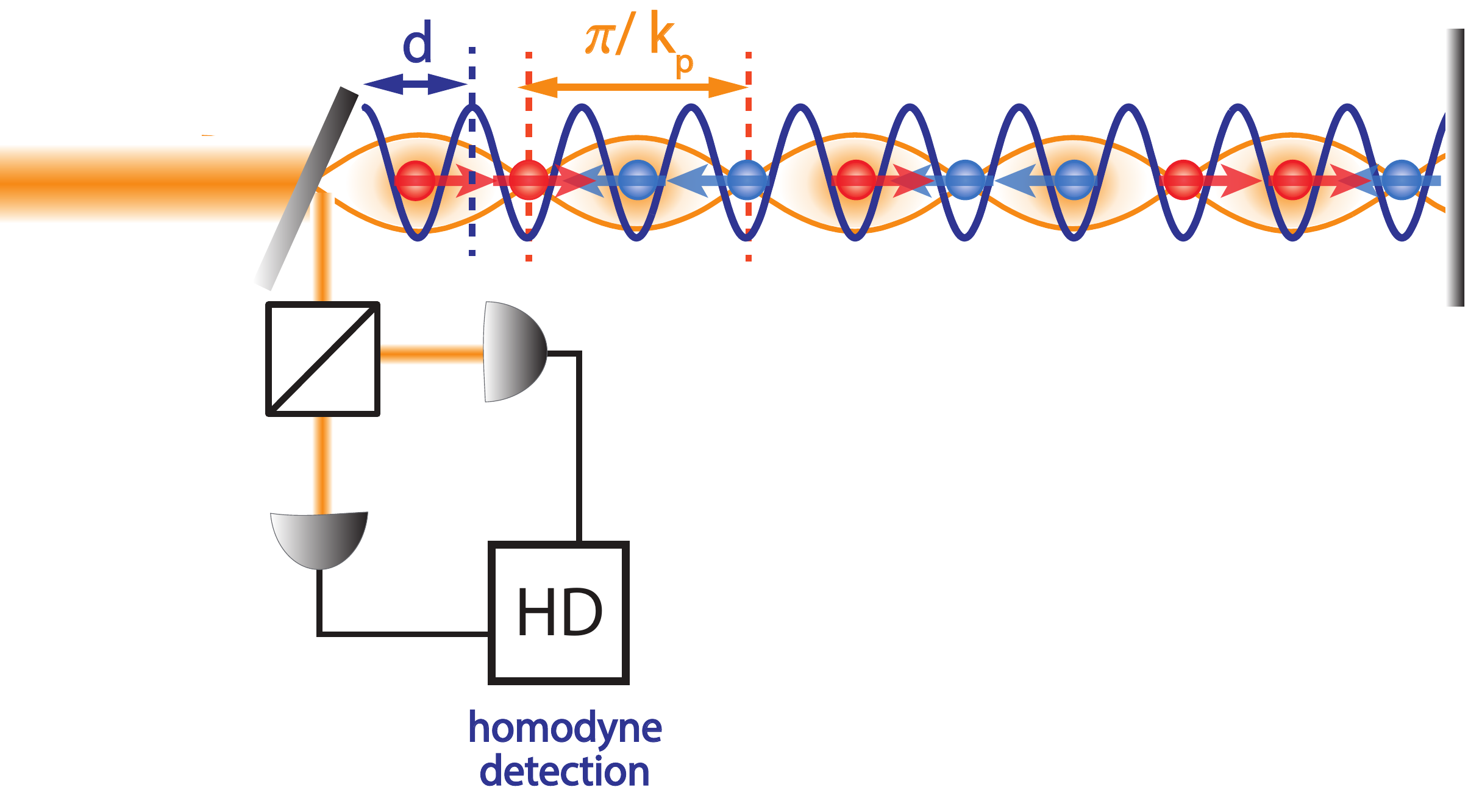}
\caption{(Color online). Schematic diagram of the proposed experimental set-up to measure the collective angular momentum imprinted on the light quadratures. The ultracold atomic sample is trapped by an optical lattice potential with wavelength $d$ (blue). An additional strong laser beam
(yellow) initially polarized in the $x$ direction is impinging on a beamsplitter. The
transmitted part of this probe is propagating through the sample and
reflected off a mirror, forming a standing wave with wavevector $k_p$. After the second pass, the
laser beam is outcoupled to a homodyne detector, where the light quadrature is measured and recorded.}
\label{Figure3}
\end{figure}

The basics of a QND Faraday spectroscopy assume a strongly linearly polarized light beam along e.g. the $x$-axis propagating on the $z$-axis and interacting off resonantly with the internal spin degree of freedom of an atomic sample. Due to the atom-photon interaction, the light polarisation is rotated by an amount that depends on the magnetic state of the sample. The light can be described by time-integrated canonical operators $X=S_{2}/\sqrt{N_{ph}}$ and $P=S_{3}/\sqrt{N_{ph}}\;$, where $S_{2(3)}$ denote the Stokes operators in the perpendicular directions of the incoming beam while $N_{ph}$ is the total number of photons of the beam. If the atomic sample is confined in an optical lattice, the light can be modulated in a standing wave configuration as schematically depicted in Fig.~\ref{Figure3}.  After the Faraday interaction has taken place, the integrated equations of motion result into\cite{Kupriyanov}

\begin{equation}
X_{\mathrm{out}} = X_{\mathrm{in}} - \frac{\kappa}{\sqrt{N}} J_z,
\end{equation}

where $X_{\mathrm{in}}$ and $X_{\mathrm{out}}$ represent, in the input-output formalism, the light quadratures before and after the Faraday interaction, and $N$ is the number of atoms, which is equal to the number of lattice sites in the ``single atom per site'' scenario. The observable $J_{z}$ corresponds to the modulated collective angular momentum along $z$-direction and is defined as:
\begin{equation}
J_z =\sum_l \cos^2\left( k_p l d\right)\; \sigma_l^z.
\end{equation}
The above sum extends on all lattice sites $l$, $k_p$ is the wave vector of the probing beam and $d$ is the inter-site distance. Finally, the light-matter coupling constant $\kappa=\sqrt{d_{o} \eta}$ depends on the optical depth of the atomic sample $d_{o}$ as well as on the spontaneous emission probability induced by the probe. Typical values of 
$\kappa$ are in the range 1-10 \cite{Hammerer14, Gajdacz13}.

As the light and atom states are initially uncorrelated, it follows that

\begin{eqnarray}
\langle X_{\mathrm{out}} \rangle &=&  - \frac{\kappa}{\sqrt{N}} \langle J_z\rangle,
\\
{\rm Var} (X_{\mathrm{out}})  &=& \frac 12 + \frac{\kappa^2}{N} {\rm Var} (J_z),
\end{eqnarray}

where we assume the incoming light beam to be in a coherent state with zero mean and variance $1/2$. For the ferromagnetic case ($J>0$),  the output signal is maximum when the wave vector of the probe beam is set to $k_p = \pi / d$, i.e. the light is not modulated.  For the antiferromagnetic case ($J<0$), since the total magnetization of the sample is zero, it is necessary to modulate the incoming beam with half of the frequency $k_p = \pi / 2d$.  

After the outcoming light quadrature $ X_{\mathrm{out}}$ has been homodyne measured, the atomic sample is projected onto a subspace of fixed $J_z$. Owing to the fact that the off resonant interaction with the light does not destroy the sample, we  further assume that after the measurement thermalization will take place on such given subspace. Since typical thermalization times for ultracold lattice gases are on the order of $ms$ and the many-body sample is stable on the time scale of seconds, the Faraday interface taking place in the $\mu s$ regime can be considered as instantaneous. Thus, the Faraday interface could be repeated several times on the same sample preserving its QND character. Finally, we remark that in order to measure the other collective operators $J_x$ and $J_y$ using the same experimental setup, one should apply an appropriate spin rotation to the atomic sample so to map $\sigma^x\to \sigma^z$ or $\sigma^y\to \sigma^z$\cite{Toth}.  
\section{Quantum thermometry for the XY model using a Faraday interface}
The quantum polarization spectroscopy technique described in the previous section grants access, a priori, to 
any order of the statistical moments of the collective atomic angular momentum, which are obtained from the values of the corresponding collective angular moments $J_{x_i}$\cite{Bark2010}. For certain phases, as for instance the paramagnetic phase, the mean value of the transverse magnetization $J_z$ is sufficient to infer the temperature of the sample. However, the mean value might vanish for other observables in an unbroken symmetry phase (e.g. the longitudinal magnetization for the thermal state in the Ising model). 
Instead, the ordering is clearly revealed when looking at the quantum fluctuations or variance of the observable. Here, for reasons that will become clearer later, we focus our study on the mean value of $J_z$ and the variance of $J_x$. The latter can be written as:

\beq
\mathrm{Var} (J_x) = \sum_{l,m} \langle \sigma_l^x \sigma_m^x\rangle-\langle \sigma_l^x\rangle\langle\sigma_m^x\rangle=\sum_{lm}\mathrm{Corr}\left(J_l^x,J_m^x\right),
\eeq
and corresponds to the sum over any two-site correlation function or, equivalently, to the magnetic structure factor at zero quasi-momentum. 
The two body correlations can be straightforwardly derived \cite{Mikeska}:
\begin{equation*}\mathrm{Corr}\left(J_l^x,J_m^x\right)=
\left\{\begin{array}{cc}
\mathrm{det}\mathrm{\textbf{G}}^r \quad &  r=l-m\neq 0 \\
1 \quad&r=l-m=0\end{array}\right.
\end{equation*}
where,
\begin{equation*}
\mathrm{\textbf{G}}^r=\left(\begin{array}{ccccccccc}
g_{-1}&&g_{-2}&&g_{-3}&&\dots&&g_{-r}\\
g_{0}&&g_{-1} &&g_{-2} &&g_{-3} &&\dots\\
g_{1}&& g_{0}&&\ddots && \ddots&&\ddots\\
\vdots&& && && &&\\
g_{r-2}&& && && &&\\
\end{array}\right) 
\end{equation*} 
And the elements are given by:
\begin{equation*}
g_j=\frac{2}{N}\sum_{k=-N/2}^{N/2-1}\left(\cos\left(\frac{2\pi}{N}kj+\theta_k\right)n_k+\sin\left(\frac{2\pi}{N}kj+\theta_k/2\right)\sin{\theta_k/2}\right)-\delta_{j,0}.
\end{equation*} 
With $n_k$ being the average number of fermionic particles in the $k$th energy level at the inverse temperature $\beta$:
\begin{equation*}
n_k=\frac{1}{1+\mathrm{e}^{\beta\epsilon_k}};
\end{equation*}
and
\begin{equation*}
\theta_k=\arctan\left(\frac{\lambda\sin\left(\frac{2\pi}{N}k\right)}{\cos\left(\frac{2\pi}{N}k\right)-\frac{h}{J}}\right).
\end{equation*}

We start by analyzing the strength of the output signal when measuring the variance of the observable associated to the order parameter, i.e. ${\rm Var} (J_x)$  for $\gamma>0$. Note that the results for $J_y$ and $\gamma<0$ are equivalent to those for $J_x$ and  $\gamma>0$. We recall that for a coherent input beam, the shot noise is ${\rm Var} (X_{\rm{in}}) = 1/2$. As expected, the variance of the operator associated to the order parameter always exceeds the variance of the angular momentum along the other two directions. Moreover, this is maximal for the Ising model ($\gamma=1$) and continuously decreases when approaching the XX model ($\gamma=0$). 

\begin{figure}[t!]
\centering
\includegraphics[width=12cm]{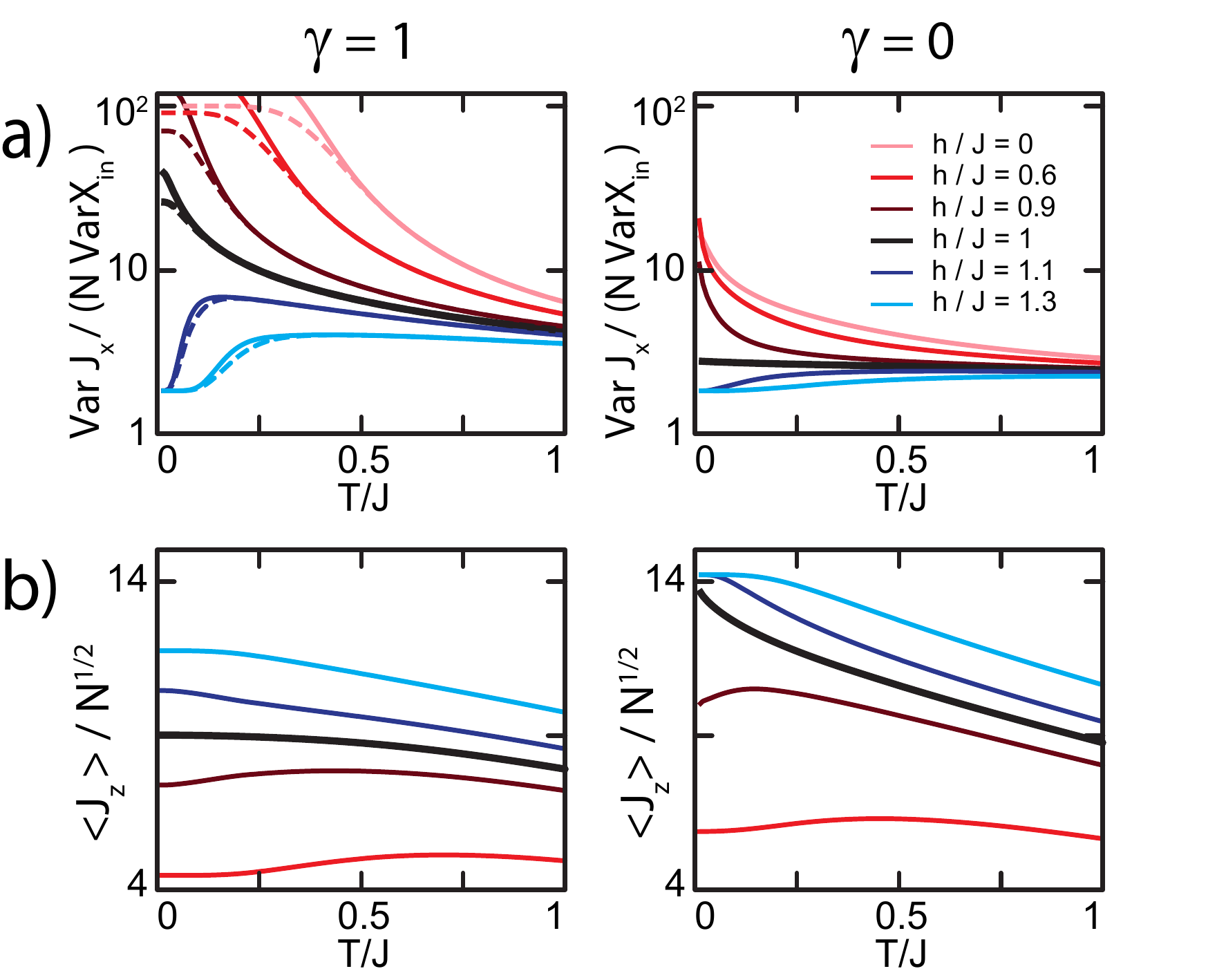}
\caption{(Color online). Output signal (assuming $\kappa=1$) as a function of ${T/J}$ for the two limiting cases $\gamma=1$ (Ising model) and $\gamma=0$ (isotropic XX model), for two observables. In red (blue/black) FM (PM/critical) phase for different values of $\mathrm{h/J}$. 
(a) $\mathrm{Var} (J_x)  / N$, normalized, for comparison, to the incoming beam shot-noise ($\mathrm{Var} (X_{\mathrm{in}}) = 1/2$). At finite $T$ and in the thermodynamic limit, $\mathrm{Var} (J_x)$ scales linearly with $N$, and the signal is always larger than $\mathrm{Var} (X_{\mathrm{in}})$. At low $T/J$, the signal decreases (increases) with $T$ in the FM (PM) phase. (b)  $\langle J_{z}\rangle/\sqrt{N}$. The mean value $\langle J_z\rangle$ scales linearly with $N$, and it shows the opposite behavior compared to (a). Solid (dashed) lines correspond to $N=200$ ($N=100$).}
\label{Figure4}
\end{figure}

A comparison between these two limiting cases ($\gamma=1$ and $\gamma=0$) is depicted in Fig.~\ref{Figure4}, where in the top panels, we display the output signal $\mathrm{Var} (J_x)  / N$ normalized by the input shot-noise $\mathrm{Var} (X_{\mathrm{in}})$ as a function of $T/J$, for different values of $h/J$ and two different system sizes $N=100$ and $N=200$. At zero temperature, and in the gapped FM phase (red line), the signal scales as $ \kappa^2 N$, whereas in the PM phase (blue line), it scales as $\kappa^2$. Strictly speaking, and since we are dealing with a 1D system, there exists no phase transition at finite temperature. This is reflected in the fact that, at any finite value of $T$, the signal in the gapped FM phase does not scale anymore as $\kappa^2 N$ as it should  be at $T=0$ where the magnetization of the pure ground state is proportional to the number of atoms, but shows a $\kappa^2$ behavior, and the signal for the two system sizes overlap. Therefore, the plateau depicted in the top panel of  Fig.~\ref{Figure4} is only a finite size effect and it disappears as the system size increases. This fact shows that for small systems ($N\le 100$), the ferromagnetic region is not useful for thermometry as the signal is constant with $T$. The results for any $\gamma\neq 0$ are qualitatively similar to those for the Ising model. Moreover, for any value of the parameters $\gamma$ and $h$ the inequality  $\mathrm{Var} (J_x) \geq 1$ is always satisfied. Therefore, if the optical depth $d_{o}$ is such that $\kappa\geq 1$, the signal of the output beam will be always greater than the input beam shot-noise. This is, however, not the case for the other two observables 
$\mathrm{Var} (J_y)$ and $\mathrm{Var} (J_z)$, which go well below the shot noise limit when approaching the XX model.

The output signal, when measuring the mean value of the $J_z$ observable ($\langle J_z  \rangle / \sqrt{N}$), is depicted in the bottom panels of Fig.~\ref{Figure4}. In contrast to the former observable, this is maximum (in absolute value) in the PM phase and it increases when approaching the $\gamma = 0$ limit.\\

In order to asses the optimality of measuring collective quantum correlations for precision thermometry, we focus on the signal-to-noise ratio $(T/\Delta T)^{2}_{\mathrm{F}}$ achievable by using the Faraday interface, and compare it
with the minimal possible error in temperature estimation, provided by the Cram\'er-Rao bound $(T/\Delta T)^2_{\mathrm{CRB}}$ (\ref{Eq:stn}). To this aim, the error performed in measuring temperature using the observable $A$ can be estimated as~\cite{braunstein_caves}
\beq
\Delta T \approx \left(\frac{\partial \langle A\rangle}{\partial T}\right)^{-1} \left(\mathrm{Var} (A)\right)^{1/2},
\eeq
\ignore{where $\Delta A$ is the standard deviation of the observable or $\left(\mathrm{Var} (A)\right)^{1/2}$.} Therefore,
\beq
\left(\frac{T}{\Delta T}\right)^2 _\mathrm{F} \approx \left(\frac{\partial \langle A \rangle}{\partial T} \right)^2\frac{T^2}{ \mathrm{Var} (A)}.
\eeq
The variance of the two observables of interest can be evaluated for the studied model. The $\mathrm{Var}(J_x^2)=\left<J_x^4\right>-\left<J_x^2\right>^2$ contains, in the first term, the sum over any four-body spin correlations ($\left<\sigma_{l_1}^x \sigma_{l_2}^x \sigma_{l_3}^x \sigma_{l_4}^x\right>$, where the subindices run over any lattice site). This can be rewritten using the Wigner-Jordan transformation as a string of fermionic operators. By using Wick's theorem it can be expanded as product of only two-body correlations (similar to what is done for the off-diagonal spin correlation functions in \cite{Lieb61}), that can be readily evaluated in the quasi-momentum representation after using the Bogoliubov transformation. The $\mathrm{Var}(J_z)$ can be directly evaluated since it only contains density-density terms.

\begin{figure}[h!]
\centering
\includegraphics[width=12cm]{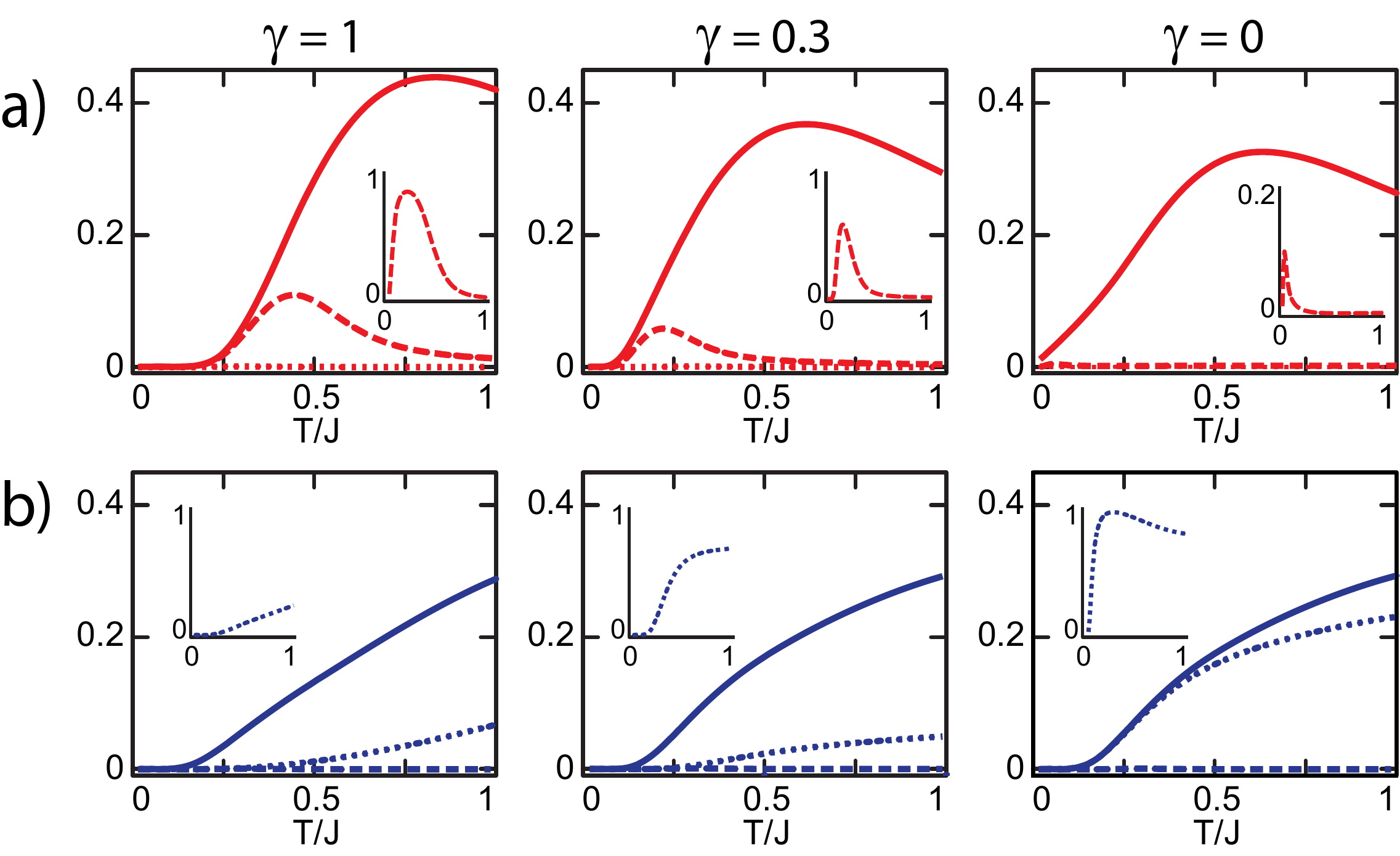}
\caption{(Color online). Comparison between the optimal signal-to-noise ratio (solid line), estimated by the Cram\'er-Rao bound, $(T/\Delta T)^{2}_\mathrm{CRB}$, and the signal-to-noise ratio obtained with the Faraday interface using $\mathrm{Var} (J_x)$ (dashed line) and $\langle J_z \rangle$ (dotted line) all normalized with the number of atoms ($N=50$ here), for different values of $\gamma$ and as a function of $T/J$. (a) $h/J=0$ (FM phase) and (b) $h/J=1.5$ (PM phase). $\mathrm{Var} (J_x)$  is optimal in the FM phase and $\gamma=1$, $\langle J_z \rangle$ is optimal in the PM phase and $\gamma =0$. The optimality of the Faraday method is depicted in the inset, where we plot the ratio between the signal-to-noise given by the Faraday and the ultimate achievable signal-to-noise given by the Cram\`{e}r-Rao bound.}
\label{Figure5}
\end{figure}

A comparison between the optimal signal-to-noise ratio  $(T/\Delta T)^2_{\mathrm{CRB}}$ (top panels), and the one obtained measuring the two observables $A=J_{x(y)}^2-\langle J_{x(y)} \rangle^2$ for $\gamma>0$ ($\gamma<0$) (middle panels)  and $A'=J_z$ (bottom panels), (all normalized by the number of atoms, $N=50$), is presented in Fig.~\ref{Figure2}, for the whole phase diagram and different temperatures. By fixing the value of $h/J$, a quantitative comparison between both signals can be performed as a function of temperature for different phases.  In Fig.~\ref{Figure5} we fix the anisotropy parameter to $\gamma=1$, $\gamma=0.3$ and $\gamma=0$, and analyze the behaviour of the FM phase ($h/J=0$) (top panel) and the PM phase ($h/J=1.5$) (bottom panel).  

These two figures show that, in general, $A$ ($A'$) performs better in the FM (PM) regions. Also, in the FM regions the signal-to-noise ratio of $A$ follows the same qualitative behavior as $(T/\Delta T)^2_{\mathrm{CRB}}$, shifting with temperature its maximum value from the multicritical points ($\gamma=0$, $|h/J|=1$) to the Ising model at $h=0$. However, it decays faster with $T/J$ than $(T/\Delta T)^2_\mathrm{CRB}$. Moreover --having in mind that the range of temperatures of interest for present experiments with ultracold  atomic gases simulating strongly correlated systems lay, approximately, in the interval $0.2<T/J<0.5$, \cite{Paredes2004,Tarruell2012}-- our results clearly show that, the Faraday spectroscopy, when reading out the observable $\mathrm{Var} (J_x)$, provides an accurate measurement of temperature in the FM phase in the Ising model, and its optimality decreases when approaching the critical XX model ($\gamma=0$). Instead, in the PM phases,  $\langle J_z \rangle$ approaches the ideal bound in the XX model for a wider temperature range.

\textit{N. B.:} The former discussion corresponds to the optimal case when the coupling $\kappa$ is very large, and the input light shot noise is negligible compared to the atomic thermal fluctuations. For more realistic values however, the result is qualitatively very similar. 

\section{Summary}
In summary, we have analyzed the suitability of QND Faraday interfaces to provide a precise estimate of the temperature of a sample of ultracold gases simulating the XY model. The Faraday interface, giving access, a priori, to any statistical moment of the collective angular momentum operators, might become optimal for this task.  Their suitability depend upon the order displayed in the strongly correlated system and the temperature range.  By borrowing concepts from quantum metrology, we have analytically derived the optimal signal-to-noise ratio for a thermal state governed by the XY Hamiltonian given by the quantum Cram{\'e}r-Rao bound, and we have compared it with the one obtained from the Faraday interface. Remarkably enough, collective atomic correlations can be considered as optimal observables for precision thermometry in the temperature range of interest in present experiments of ultracold lattice gases simulating strongly correlated systems. Our results hold for the XY model, but it remains to be analyzed if the method can also be optimal for other quantum spin models, either integrable or not. 

\ack We are particularly thankful to John Calsamiglia and Luis Correa for their valuable comments, and to Lorenzo del Re for useful discussions. Financial support from EU Collaborative
Project TherMiQ (Grant Agreement 618074), Spanish MINECO (FIS2008-01236), European Regional Development Fund, Generalitat de Catalunya (Grant No. SGR2014-9), the UK EPSRC (EP/L005026/1, EP/K029371/1), the John Templeton Foundation (grant ID 43467), the COST Action MP1209 is acknowledged.

\section*{References}
\bibliographystyle{iopart-num}
\bibliography{biblio}
\end{document}